\documentclass[preprint2]{aastex} 
\slugcomment{to submit to ApJL}
\usepackage{natbib}

\shorttitle{On the dependence of SNe~Ia luminosities on their host galaxies metallicity}
\usepackage{graphicx} 
\usepackage[dvips]{epsfig}
\usepackage{txfonts}
\usepackage{subfigure}
\usepackage{rotating}

\DeclareRobustCommand{\ion}[2]{%
\relax\ifmmode
\ifx\testbx\f@series
{\mathbf{#1\,\mathsc{#2}}}\else
{\mathrm{#1\,\mathsc{#2}}}\fi
\else\textup{#1\,{\mdseries\textsc{#2}}}%
\fi}

\newcommand{\HII}{\ion{H}{ii}}

\newcommand{\OHb}{[\ion{O}{iii}]\,$\lambda$5007/H$\beta$}
\newcommand{\NHa}{[\ion{N}{ii}]\,$\lambda$6583/H$\alpha$}

\begin{document}

\title{On the dependence of the type Ia SNe luminosities\\ 
on the metallicity of their host galaxies}
\author{Manuel E. Moreno-Raya}
\affil{Dpto.de Investigaci\'{o}n B\'{a}sica, C.I.E.M.A.T.,
Avda. Complutense 40, 28040 Madrid, Spain }
\email{manuelemilio.moreno@ciemat.es}
\author{Mercedes Moll\'{a}}
\affil{Dpto.de Investigaci\'{o}n B\'{a}sica, C.I.E.M.A.T.,
Avda. Complutense 40, 28040 Madrid, Spain }
\author{\'Angel R. L\'{o}pez-S\'{a}nchez}
\affil{Australian Astronomical Observatory, PO Box 915, North Ryde, NSW 1670, 
Australia;\\ Department of Physics and Astronomy, Macquarie
University, NSW 2109, Australia}
\author{Llu{\'\i}s Galbany}
\affil{Millennium Institute of Astrophysics MAS, Nuncio Monse\~{n}or S\'{o}tero Sanz 100, Providencia,7500011 Santiago, Chile;\\
 Departamento de Astronom{\'\i}a, Universidad de Chile, Camino El Observatorio 1515, Las Condes, Casilla 36-D, Santiago, Chile. }
\author{Jos\'{e} Manuel V\'{\i}lchez}
\affil{Instituto de Astrof\'{\i}sica de Andaluc\'{\i}a (CSIC), 
Apdo. 3004, 18080 Granada, Spain}
\author{Aurelio Carnero Rosell}
\affil{Observat\'{o}rio Nacional, and LIneA Laborat\'{o}rio Interinstitucional de e-Astronomia, \\
Rua Gal. Jos\'{e} Cristino 77 Rio de Janeiro, RJ 20921-400, Brazil; }
\author{Inmaculada Dom\'{\i}nguez}
\affil{Departamento de F\'{\i}sica Te\'{o}rica y del Cosmos, Universidad de Granada, 18071 Granada, Spain}
\vspace{-1cm}
\begin{abstract}
The metallicity of the progenitor system producing a type~Ia
supernova (SN~Ia) could play a role in its maximum luminosity,
as suggested by theoretical predictions. We present an observational study to
investigate if such a relationship there exists. Using
the 4.2m WHT we have obtained intermediate-resolution spectroscopy data of a
sample of 28 local galaxies hosting SNe~Ia, for which distances have
been derived using methods independent to those based on the own
SN~Ia parameters. From the emission lines observed in their optical spectrum,
we derived the gas-phase oxygen abundance in the region where each SN Ia exploded. Our data 
show a trend, with a 80\% of chance not to be due to random fluctuation, between SNe~Ia absolute magnitudes and the
oxygen abundances of the host galaxies, in the sense that luminosities
tend to be higher for galaxies with lower 
metallicities. This result seems like to be in agreement with both the theoretically expected behavior, 
and with other observational results. This dependence $M_{B}$-Z might induce to
systematic errors when is not considered in deriving SNe~Ia luminosities and
then using them to derive cosmological distances.
\end{abstract}


\section{INTRODUCTION}

The Supernova Cosmology is based on the well-known Hubble Diagram (HD) in
which distances of SNe~Ia are represented as a
function of their redshifts, $z$, usually determined
with high accuracy from SNe~Ia or host galaxies spectra. 
Distances are derived by means of the distance modulus, $\mu=m-M$.
\citet{phillips93,hamuy96a,hamuy96b}, and \citet{phillips99}
found a correlation between the properties of the SN~Ia light
curve (LC) and the absolute magnitude in its maximum for $B$, $V$ and $I$
bands. Therefore, under the assumption that SNe~Ia are standard-calibrated candles,
their absolute magnitude $M$ can be obtained from empirical
calibrations based on their observed LCs.
Thus, the SNe~Ia-based cosmology projects discovered the Universe is in accelerated expansion \citep{perl99,riess98}. 

However, the SNe~Ia methodology is 
calibrated on local objects, whose host galaxies probably
share almost solar abundances\footnote{Here we use the terms
metallicity or total abundance in metals, Z, (being X+Y+Z=1 in mass),
and oxygen abundances indistinctly, assuming $\log(\rm
Z/Z_{\odot})=\log({\rm O/H})-\log({\rm O/H})_{\odot}$, $12+\log({\rm
O/H})_{\odot}=8.69$, and $Z_{\odot}=0.016$ being the solar values
\citep{asp09}.}. But chemical abundances change with redshift
\citep[e.g.][and references therein]{Lara09} due to chemical evolution of galaxies, therefore
the LC calibration might not be the same for chemical abundances which differ very much from solar values. 
This possible metallicity dependence of the SNe~Ia luminosity has been neglected 
but it might play a role in accurately determining the
distances to cosmological SNe~Ia. Since the number of SNe~Ia detections
 will extraordinarily increase in the forthcoming surveys
(DES\footnote{http://www.darkenergysurvey.org}, LSST\footnote{http://www.lsst.org/lsst/}), statistical errors
will decrease while systematic errors will begin to dominate, limiting
the precision of SNe~Ia as indicators of extragalactic
distances. The metallicity dependence may be 
one source of systematic errors when using these techniques,
being therefore important to quantify its effect. 
 
A dependence of the maximum luminosity of SNe~Ia on the initial
metallicity of their progenitors is theoretically predicted. 
SNe~Ia are thought to be thermonuclear explosions of carbon-oxygen white
dwarfs (WD) in binary systems, \citep{whelan73,iben84,webbink84}.
The WD approaches the critical Chandrasekhar mass by accretion from
its companion. The maximum luminosity of
a SN~Ia depends on the amount of $^{56}$Ni synthesized during the
explosion \citep{arnett82}:
\begin{equation}
L\propto M(^{56}\rm Ni)~~erg\,s^{-1},
\label{bravo1}
\end{equation}
Assuming the mass of the exploding WD is constant, the parameter
who leads the relation between the LC parameters and its maximum
magnitude is the outer layer opacity of the ejected material
\citep{hof96}, which depends on temperature and, thus, on the
heating due to the radioactive $\rm ^{56}Ni$ decay. 
The neutron excess in the exploding WD, which controls
the radioactive ($\rm ^{56}Ni$) to non-radioactive (Fe-peak elements)
abundance ratio, depends directly on the initial metallicity of the progenitor. 
Therefore the maximum luminosity of the SN~Ia explosion depends on
the initial abundances of C, N, O, and Fe of the WD progenitor \citep{timmes03,tra05,pod06}. 
\citet{timmes03} found a dependence on Z as $1-0.057{\rm Z/Z_{\sun}}$.
\cite{bravo10} computed a series of SNe~Ia explosions, finding a stronger dependence on Z:
\begin{equation} 
M(^{56}{\rm Ni}) \sim f({\rm Z})=1-0.075\frac{\rm Z}{\rm Z_{\odot}}
\label{bravo2}
\end{equation} 

They also explored the dependence of the explosion on the local chemical composition, 
finding a non-linear relation : 
\begin{equation} 
M(^{56}Ni) \sim f({\rm Z})=1-0.18\frac{\rm Z}{\rm Z_{\odot}}\Bigg(1-0.10\frac{\rm Z}{\rm Z_{\odot}}\Bigg)
\label{bravo3}
\end{equation}
Since the luminosity decreases with Z increasing,
SNe~Ia located in galaxies with $\rm Z>Z_{\odot}$ {\it might be dimmer} than expected as compared to those
with $\rm Z \le Z_{\odot}$ 

The dependence of SNe~Ia luminosities on the metallicity was
studied by \citet{gal05}, who estimated oxygen elemental
abundances by using host-galaxies emission lines, finding most metal-rich galaxies have the faintest SNe~Ia. They based their
conclusion on the analysis of the HD residuals, implying they used SNe~Ia 
to extract the magnitudes. \citet{gal08} analyzed spectral absorption indices in early-type galaxies, 
using theoretical evolutive synthesis models (still not very precise, S{\'a}nchez et al. in prep.), also
finding a trend between SNe~Ia magnitudes and the metallicity of their stellar populations. 
These results are in agreement with theoretical predictions. 

Other dependences of SNe Ia magnitudes have been found
studying the correlations between the residuals in the HD, as $\mu-\mu_{fit}$, and the host
galaxy characteristics. SNe Ia in massive galaxies 
result brighter than spected {\bf after} correcting for their LC widths and colors 
\citep[see][and references therein for more details]{howell09,neill09,sull10,lam10,childress13,pan14,betoule14,moreno15}.
Dividing the galaxy mass range into two groups, the $M_B$ of SNe~Ia shows a step of $\sim$ 0.07-0.10\,mag\,dex$^{-1}$ 
in the residuals plot \citep[see ][mainly Table 2, where observational trends from different authors are compiled]{childress13}
between both bins. 

Following \citet{rigault13} and \citet{galbany14}, our aim is to perform a systematic study to determine if SNe~Ia luminosities depend on
the {\it local} elemental abundances of their host galaxies. We built a sample of nearby galaxies hosting SNe~Ia,
selecting objects {\bf for which distances were determined using methods different to those based on SN~Ia}. 
We then conducted intermediate-resolution long-slit spectroscopic observations of the sample to estimate the oxygen gas-phase abundances
and, when possible, derive the {\bf local} metallicity
around the region where SNe~Ia exploded.
With that we directly check the possible luminosity-metallicity relationship.

\section{OBSERVATIONS AND OXYGEN ABUNDANCES}
\label{obs}
We have observations for 28 local galaxies hosting SNe~Ia with the 4.2m William
Herschel Telescope (WHT) at El Roque de Los Muchachos Observatory,
La Palma, Spain, in two campaigns in December 2011 (9 galaxies) and January 2014 (19 galaxies). We observed more galaxies but they did not show emission lines with sufficient S/N ratio to measure their intensities. Therefore by construction, our sample is biased to star-forming galaxies.
We used the two arms of the ISIS spectrograph, covering from 3600 to 5200~\AA\ in the blue and
from 5850 to 7900\,\AA\ in the red, with 0.45~\AA/pix and 0.49~\AA/pix, respectively. Galaxies were chosen
from \citet{neill09}, selecting objects not in the Hubble flow ($z \leq 0.02$) 
and for which distances not based on SNe~Ia data are
available. Table~\ref{sample} compiles the details of the observed sample.
89 \ion{H}{ii} regions were analyzed. We have many galaxies with oxygen abundances estimates for several regions, 
for which we determined a metallicity radial gradient \citep[]{moreno15}. 

\begin{table*}[ht!]
\footnotesize
\caption{Sample of observed galaxies and their associated SNe~Ia. 
Galaxy names in column 1, galaxy center coordinates RA and DEC in columns 2 and 3, and hosted SNe~Ia in column 4, host galaxies magnitudes, $M_{B}$, 
in column 5, and the corresponding redshift $z$ in column 6. Distance indicator and SN Ia class, as reddened (R) , normal (N) or subluminous (S), 
are shown in columns 7 and 8., and LC fitters in column 9.}
\centering
\begin{tabular}{l@{\hspace{10pt}} c@{\hspace{10pt}} c@{\hspace{10pt}}  l@{\hspace{10pt}} c@{\hspace{10pt}}c@{\hspace{15pt}}c@{\hspace{15pt}}c@{\hspace{15pt}}c}
\noalign{\smallskip}
\hline
{Object} & {RA} &{DEC} &{SN~Ia} &{$\rm M_{B}$} &{$z$}&{Distance indicator$^{*}$}&{SN~Ia class}&{LC fitter}\\

 \hline 
\noalign{\smallskip}
MGC~021602 & 06 04 34.9 &  -12 37 29  &   2003kf &  -19.86  &   0.007388  &  TF    &   N    & SALT2 \\
NGC~0105   & 00 25 16.6 &  +12 53 22  &   1997cw &  -20.98  &   0.017646  &  SN~Ia    &   R    & SALT2 \\ 
NGC~1275   & 03 19 48.1 &  +41 30 42  &   2005mz &  -22.65  &   0.017559  &   TF   &   S    & SALT2 \\ 
NGC~1309   & 03 22 06.5 &  -15 24 00  &   2002fk &  -20.57  &   0.007125  &  CEPH \& TF    &    N   & SALT2 \\
NGC~2935   & 09 36 44.8 &  -21 07 41  &   1996Z  &  -20.69  &   0.007575  &   TF   &  R     & SALT2 \\  
NGC~3021   & 09 50 57.1 &  +33 33 13  &   1995al &  -19.86  &   0.005140  &  CEPH \& TF    &    N   & SALT2 \\  
M~82             & 09 55 52.7 &  +69 40 46  &   2014J  &  -20.13  &   0.000677  &  PNLF    &   N    & ----- \\  
NGC~3147   & 10 16 53.6 &  +73 24 03  &   1997bq &  -22.22  &   0.009346  &   TF   &  N     & SALT2 \\  
NGC~3169   & 10 14 15.0 &  +03 27 58  &   2003cg &  -20.42  &   0.004130  &   TF   &    R   & SALT2 \\  
NGC~3368   & 10 46 45.7 &  +11 49 12  &   1998bu &  -20.96  &   0.002992  &  CEPH \& TF    &    R   & MLCS2k2 \\  
NGC~3370   & 10 47 04.0 &  +17 16 25  &   1994ae &  -19.77  &   0.004266  &   CEPH \& TF   &    N   & MLCS2k2 \\  
NGC~3672   & 11 25 02.5 &  -09 47 43  &   2007bm &  -20.63  &   0.006211  &   TF   &    R   & SALT2 \\  
NGC~3982   & 11 56 28.1 &  +55 07 31  &   1998aq &  -19.91  &   0.003699  &  CEPH \& TF    &    N   & MLCS2k2 \\  
NGC~4321   & 12 22 54.8 &  +15 49 19  &   2006X  &  -22.13  &   0.005240  &  CEPH \& TF    &    R   & SALT2 \\  
NGC~4501   & 12 31 59.1 &  +14 25 13  &   1999cl &  -23.13  &   0.007609  &   TF   &    R   & SALT2 \\  
NGC~4527   & 12 34 08.4 &  +02 39 13  &   1991T  &  -21.55  &   0.005791  &   CEPH \& TF   &   N    & MLCS2k2 \\  
NGC~4536   & 12 34 27.0 &  +02 11 17  &   1981B  &  -21.85  &   0.006031  &   CEPH \& TF   &   N    & MLCS2k2 \\  
NGC~4639   & 12 42 52.4 &  +13 15 27  &   1990N  &  -19.24  &   0.003395  &  CEPH \& TF    &   N    & MLCS2k2 \\  
NGC~5005   & 13 10 56.2 &  +37 03 33  &   1996ai &  -21.48  &   0.003156  &   TF   &    R   & SALT2 \\   
NGC~5468   & 14 06 34.9 &  -05 27 11  &   1999cp &  -20.33  &   0.009480  &   TF   &   N    & SALT2 \\   
NGC~5584   & 14 22 23.8 &  -00 23 16  &   2007af &  -19.69  &   0.005464  &   CEPH \& TF   &    N   & SALT2 \\  
UGC~272    & 00 27 49.7 &  -01 12 00  &   2005hk &  -19.42  &   0.012993  &  TF    &  S     & MLCS2k2 \\   
UGC~3218   & 05 00 43.7 &  +62 14 39  &   2006le &  -22.17  &   0.017432  &  TF    &  N     & SALT2 \\  
UGC~3576   & 06 53 07.0 &  +50 02 03  &   1998ec &  -20.98  &   0.019900  &   TF   &  N     & SALT2 \\  
UGC~3845   & 07 26 42.7 &  +47 05 38  &   1997do &  -19.95  &   0.010120  &  TF    &   N    & SALT2 \\  
UGC~4195   & 08 05 06.9 &  +66 46 59  &   2000ce &  -20.71  &   0.016305  &   TF   &   R    & SALT2 \\     
UGC~9391   & 14 34 37.0 &  +59 20 16  &   2003du &  -17.85  &   0.006384  &  TF    &   N    & SALT2 \\  
UGCA~17    & 01 26 14.4 &  -06 05 39  &   1998dm &  -19.86  &   0.006535  &  TF    &   R    & ----- \\
\hline
\end{tabular}

$^{*}$TF=Tully-Fisher; CEPH=Cepheids, PNLF=Planetary Nebulae Luminosity Function; SN~Ia=Supernova
Ia
\label{sample}
\end{table*}

Only 63 \ion{H}{ii} regions could be classified according to the
classical diagnostic diagrams \citep{BPT81}, since the [\ion{O}{iii}]~$\lambda$5007 
was not detected in 26 of them. Other 56 were
unambiguously classified as pure star-forming regions; 4 regions are 
within the \citet{kew01} and \citet{kau03} lines in the
\OHb\ vs. \NHa\ diagrams, and considered composite objects, 
still included in our analysis; 3 regions were
classified as AGNs and are not longer considered in our analysis. 
We then define our {\it subsample} with 60 star-forming +
composite objects.

\begin{figure}[t!]
\centering
\includegraphics[width=0.47\textwidth,angle=0]{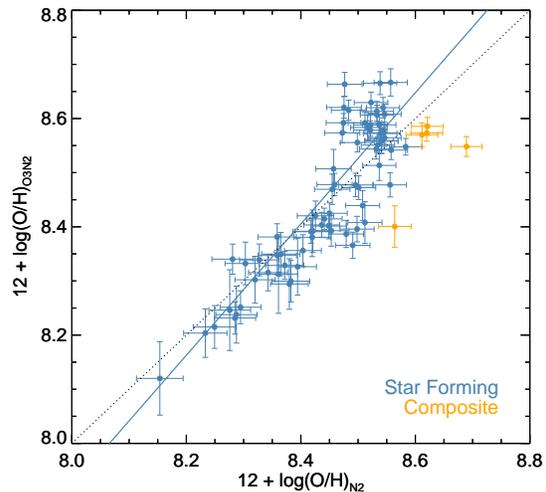}
\caption{Oxygen abundances estimated with the N2 parameter, 
compared with those obtained using the O3N2 parameter, 
for the subsample. Blue and orange dots correspond to star-forming and
composite regions, respectively. The dashed line is the identity line,
while the solid line is a linear fit to the data.
\label{OH-MAR}}
\end{figure}

Values obtained for the 26 objects for which
[\ion{O}{iii}]~$\lambda$5007 is not available 
are below the low-limit to be defined as AGNs in the \NHa\ distribution.
Hence, we consider these 26 objects as \HII\ regions too, and we will use their
emission lines to estimate their oxygen
abundances. Our final {\it sample} has a total of 56+4+26=86 \HII\
regions.

As the faint auroral lines used to compute the electron temperature of
the ionized gas (e.g., [\ion{O}{iii}]~$\lambda$4363) are not detected
in any case, we use empirical calibrations \citep{arls12}
to estimate the oxygen abundance. We use $N2$ and
$O3N2$ parameters \citep{allo79}, and the calibrations by \citet{mar13}\footnote{With this calibration (improved using CALIFA data), is difficult to obtain abundances over 8.7 dex. Photoionization models \citep{kew01} overestimate direct abundances around 0.3-0.5 dex \citep{arls12}.}. With the
parameter $O3N2$ we obtain OH$_{O3N2}=12+\log{(O/H)}_{O3N2}$ 
for the {\it subsample}, and with $N2$, we have 
OH$_{N2}=12+\log{(O/H)}_{N2}$ for the full {\it sample}.
Figure~\ref{OH-MAR} compares OH$_{O3N2}$ and OH$_{N2}$ 
for the subsample. A least-squares linear fit yields:
\begin{equation} 
OH_{O3N2} = 1.15 (\pm 0.09) -1.23(\pm 0.77)\times OH_{N2},
\label{MARfit}
\end{equation} 
(correlation coefficient $r$=0.88), expression used to
convert OH$_{N2}$ abundances to OH$_{O3N2}$ for
those 26 regions lacking of $O3N2$ data.

Once OH$_{O3N2}$ is determined, we assign an oxygen abundance to the region within
each galaxy where its SN~Ia was located. For this we use this
procedure:
\begin{enumerate}
\item[a)] In 21 galaxies where several \ion{H}{ii} were detected, we
estimate a radial oxygen gradient and then we use it to compute
the oxygen abundance which corresponds to the projected galactocentric distance
at which the SN~Ia exploded.
\item[b)] For 7 galaxies for which the previous method cannot be applied (i.e., few \ion{H}{\sc ii} regions or unreal gradient), we just select the abundance
corresponding to the closest region to the SN~Ia, being the typical distances $\sim$2-3\,kpc.
\end{enumerate} 

A typical difference of $\sim 0.05$ dex, smaller that the oxygen abundance error ($~0.10$ dex), is found between metallicities derived using gradient or closest region methods,
 implying their results agree. We finally got the local oxygen abundance for the whole sample of 28 SNe~Ia.
 
\begin{figure*}[t]
\includegraphics[width=0.5\textwidth,angle=0]{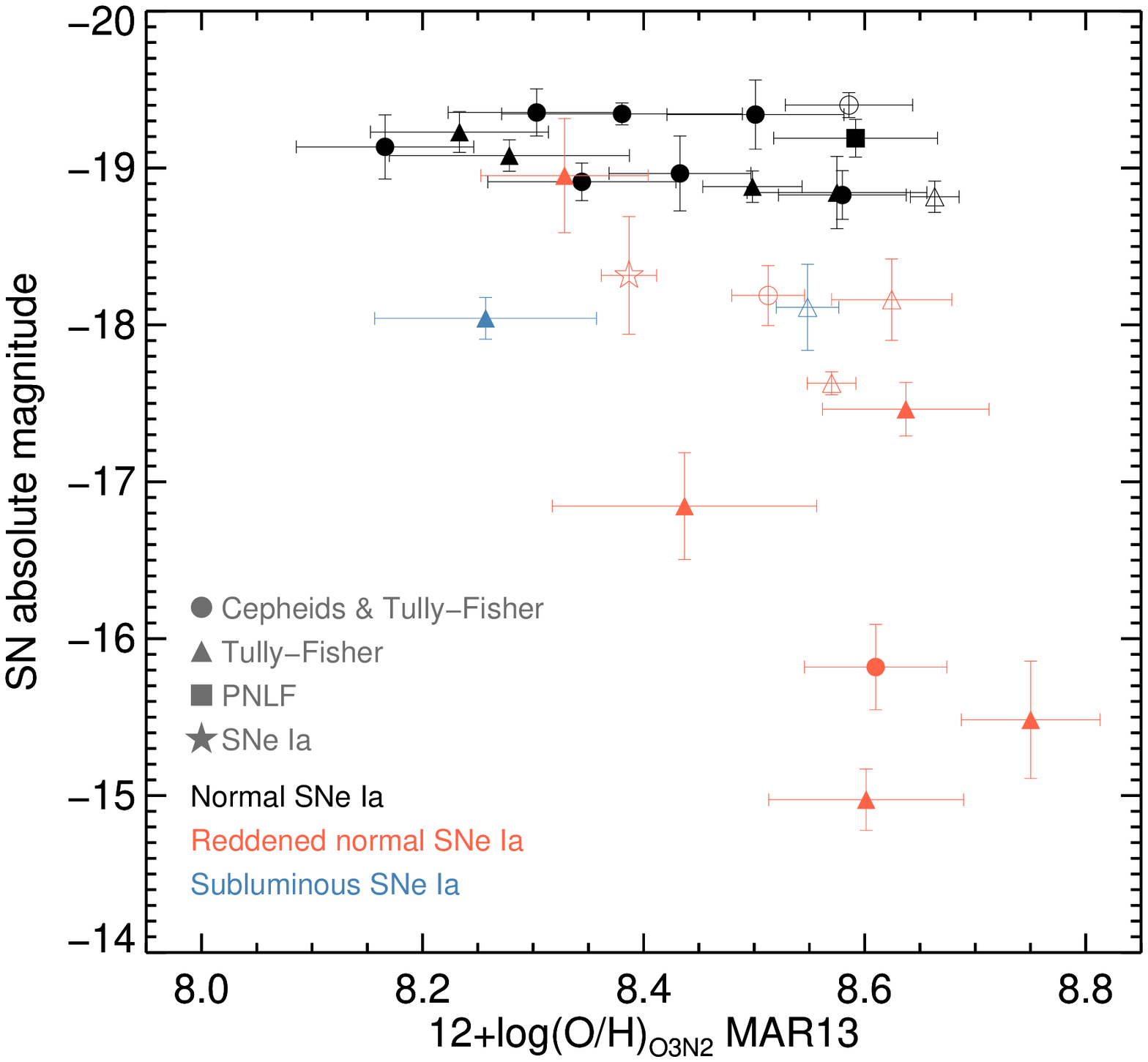}
\includegraphics[width=0.5\textwidth,angle=0]{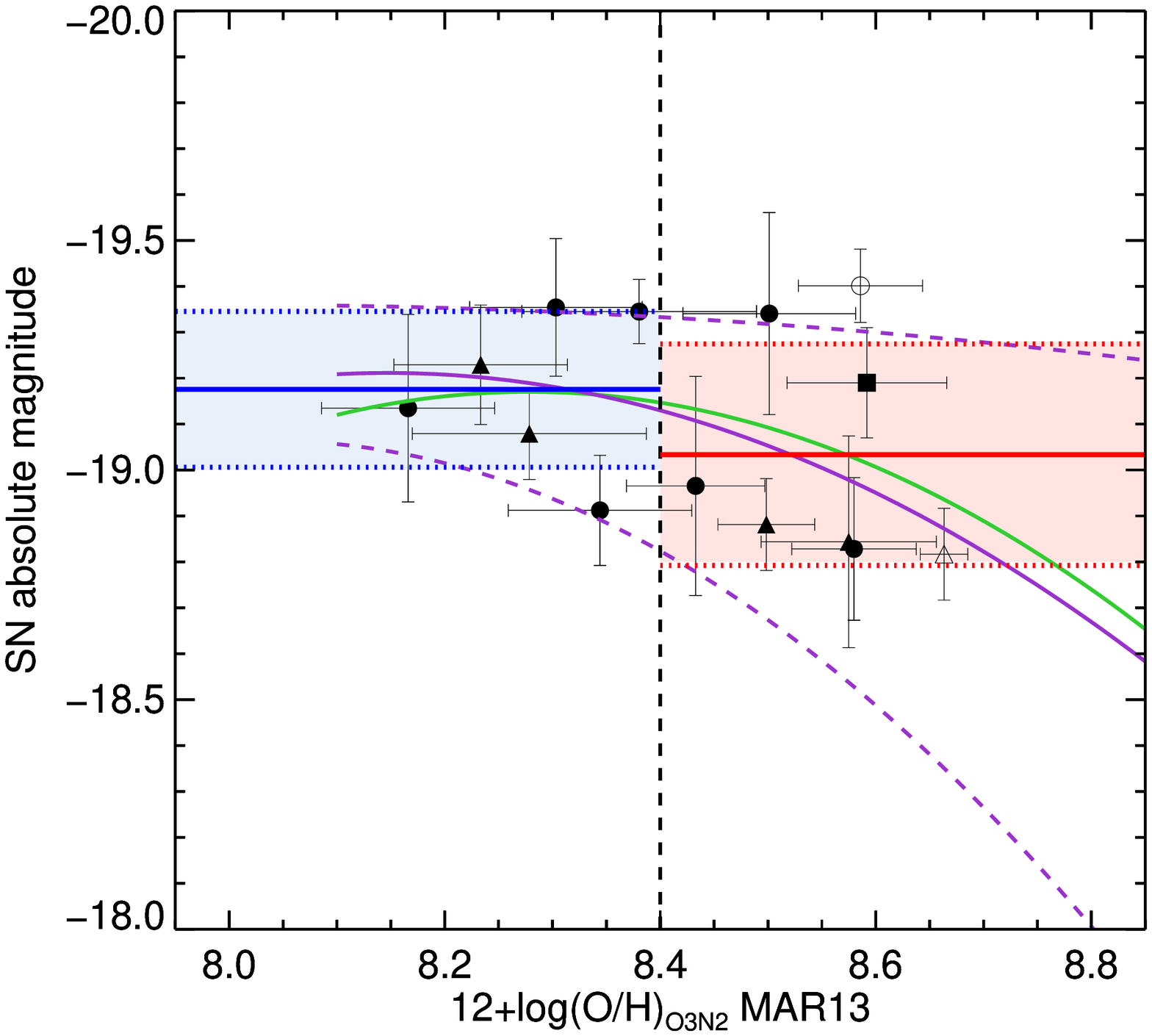}
\caption{Left: SNe~Ia absolute magnitudes,
$M_{B}$, as a function of oxygen abundances, $\rm OH_{O3N2}$.
Full and open symbols indicate objects with
abundances estimated by procedure a) and b), respectively. Triangles, squares
and circles indicate the method used to derive distances: Tully-Fisher, PNe Luminosity Function
or Tully-Fisher and Cepheids, respectively. The color indicates the type of
SN~Ia: normal (black), reddened (red), or subluminous (blue).
Right: The same plot considering only normal SNe~Ia. 
The green solid line is a second-order polynomial fit to the
data. The purple solid line is a second order polynomial fit to
four metallicity bins (see text). Short-dashed purple lines
represent the 1-$\sigma$ uncertainty to this fit. Blue and red
horizontal solid lines provide the averaged value in the low and high
metallicity regime, respectively, with their 1-$\sigma$ uncertainty 
shown with the pale blue and red areas.}
\label{mb-oh}
\end{figure*}

\section{RESULTS}
\label{results}

\subsection{Estimation of distances and magnitudes} 

The distance, $D$, to the galaxies was obtained using the NASA Extragalactic
Database, NED\footnote{\it http://ned.ipac.caltech.edu/}, selecting
those independent of the SNe~Ia method. The apparent magnitudes,
$m_{B}$, of SNe~Ia in the maximum of their LCs are
from \citet{neill09}, except for SN2014J taken directly as $M_{B}$ from \citet{marion15}.
We have corrected from the Milky Way extinction\footnote{Following \citet{neill09} the host galaxy extinction is not accounted for 
the apparent magnitudes estimation.} using the NED values for B-band, $A_{B}$:
$m_{B,ext}=m_{B}-A_{B}$. The absolute magnitudes of SNe~Ia were computed applying the usual expression:
$M_{B}=m_{B,ext}- 5\log{D} + 5$, ($D$ in pc). No standardization technique has been used to obtain the magnitudes.

\subsection{The relation between the SN~Ia luminosity and the metallicity}

Left panel of Fig.~\ref{mb-oh} plots SNe~Ia absolute magnitudes, $M_{B}$, as a function of the oxygen abundance, $\rm OH_{O3N2}$. This plot indicates SNe~Ia located in metal-rich galaxies are less luminous that the ones
in metal-poor galaxies. Right panel of Fig.~\ref{mb-oh} shows the same 
only considering normal objects \citep[eliminating reddened and
sub-luminous SNe~Ia as explained in][]{moreno15}, where we have fit
a second order polynomial function. We have studied the goodness of this last fit via a \mbox{$\chi^{2}$}
test. As errors in both magnitudes and metallicities are relatively
large\footnote{The average uncertainties of $M_B$ and $\rm OH_{O3N2}$
are $\pm$0.15~mag and $\pm$0.08~dex, respectively, while 
their values ranges are 0.85~mag and 0.50~dex, respectively.}, we have also
considered millions of random variations of the values following a
Gaussian distribution of the uncertainties in each axis. In each
iteration we fit a second-order polynomial to the data and derive the
\mbox{$\chi^{2}$} of the fit. We sought the minimum values of these values,
which confirm the relationship is satisfied with around a 80\% of probability.

We have also averaged our data in four metallicity bins: $x<8.2$, $8.2<x<8.4$,
$8.4<x<8.6$ and $x>8.6$, being $x=\rm OH_{O3N2}$. The purple
continuous line plotted in the right panel of Fig.~\ref{mb-oh} is a
second order polynomial fit to the average value obtained for these 
bins. This fit matches well with that obtained considering all data. 
Dividing the abundances into a low-metallicity, $\rm OH_{O3N2}>$8.4, and a
high-metallicity, $\rm OH_{O3N2}<$8.4, regime, we find a difference of
{\bf $0.14 \pm 0.10$}~mag in $M_B$ (blue and red horizontal lines in the right panel of
Fig.~\ref{mb-oh}), with high (low) metallicity galaxies hosting less
(more) luminous SNe~Ia.

A shift in the magnitudes as due to the metallicity effect over the
SNe~Ia luminosity is theoretically expected. Considering
Eqs.~\ref{bravo1} to \ref{bravo3}, $L \propto f({\rm Z})$, and
$M_{bol}=-2.5\log {L} \sim -2.5\log{[f({\rm Z})]}$. Assuming that
these equations are also valid in the $B$-band, the difference between
the $M_{B}$ of a system with solar abundance and the corresponding
$M_{B,{\rm Z}}$ for any other value of Z can be computed using the
functions $f({\rm Z})$ given in Eq.~\ref{bravo1} and \ref{bravo2}.
Thus, we get a metallicity-dependent magnitude:
\begin{equation} 
M_{B}({\rm Z})=M_{B,{\rm Z}_{\sun}}+\Delta M_{B}({\rm Z})~{\rm mag},
\label{nueva}
\end{equation}
where the term $\Delta M_{B}$(Z) produces a shift in the expected value $M_{B}$
which corresponds to Z$_{\odot}$.

By using Eqs.~\ref{bravo2} and \ref{bravo3}, we derive
these two metallicity-dependent relationships:
\begin{equation}
\Delta M_{B}({\rm Z})=-2.5 \log{ \Bigg( 1-0.075\frac{{\rm Z}}{{\rm Z}_{\odot}}\Bigg)}- 0.0846 \,{\rm mag},
\label{deltaZ1}
\end{equation}
and
\begin{eqnarray}
\nonumber 
\Delta M_{B}({\rm Z}) = -2.5 \log{ \Bigg[1-0.18\frac{{\rm Z}}{{\rm Z}_{\odot}} 
 \Bigg(1-0.10\frac{{\rm Z}}{{\rm Z}_{\odot}}\Bigg)\Bigg]}\\ 
 -0.191\,{\rm mag.}
\label{deltaZ2}
\end{eqnarray}

The terms 0.0846 and 0.191\,mag have been introduced to normalize these
equations to satisfy $\Delta M_{B}({\rm Z_{\odot}})=0$. Actually,
these values represent the magnitude difference between objects with
Z=0 and Z=$\rm Z_{\odot}$. As we see, the order of magnitude of these
variations, $\sim 0.10-0.20$\,mag, agrees with our observational difference of
{\bf $\sim 0.14 \pm 0.10$}~mag between the low- and high-metallicity regimes. 

\subsection{The effect of the color in the $M_B$--Z relation}

\begin{figure}[t]
\includegraphics[width=0.45\textwidth,angle=0]{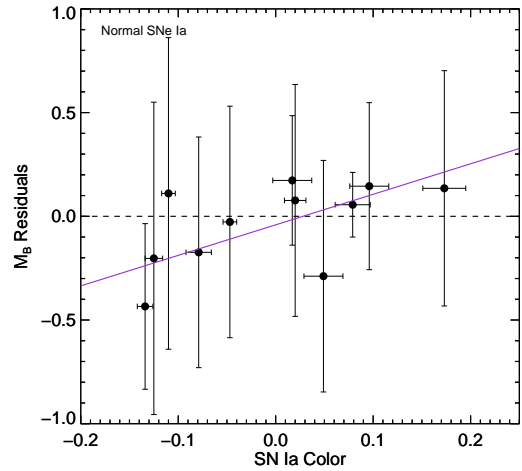}
\caption{$M_B$ residuals after subtracting the fit obtained in Fig.2a as a function of the SNe~Ia color.
The dashed line marks the zero residuals. The solid purple line is the least squares straight line to the dots.}
\label{res-color}
\end{figure}

This probable metallicity dependence of the
luminosity of the SN~Ia could be attributed to the {\it color}
correction, a term already included in the cosmological methods
to estimate the distance modulus (and implicitly in the determination
of the $M_{B}$ of each SN~Ia). Actually, the SNe~Ia color shows
a dependence on the oxygen abundance and there is also a good correlation
between SNe~Ia magnitudes and their colors \citep[see both in][]{moreno15}. 
However, when we plot Figure~\ref{res-color} with the $M_B$ residuals of the fit found in Fig.\ref{mb-oh}b, 
as a function of the SNe~Ia color, we found, as expected, that there is not a strong correlation,
implying that this parameter does not affect very much in the determination of $M_{B}$ for these not reddened
objects. A linear fit applied to these points results in:
\begin{equation} 
\Delta M_{B}= -0.04\, (\pm0.12) + 1.47\, (\pm 1.42)\times Color
\end{equation} 
This fit has a correlation coefficient $r$=0.5, and considering the errors of the fit parameters, a 
no correlation is equally valid or statistically significant.
Therefore, in agreement with \citet{childress13}, the color dependence is not sufficient to explain the correlation 
seen in Fig.~\ref{mb-oh}, and a metallicity dependence on $M_B$ is still left.

We then conclude that a correlation between the absolute magnitude of
SNe~Ia, $M_{B}$, and their host-galaxy metallicity
seems likely. Metal-rich galaxies host less bright SNe~Ia than the metal-poor galaxies. This
dependence is not included in the term of color when deriving
the distance modulus using SNe~Ia data.

\subsection{Implications of the $M_B$--Z relation}

If the $M_B$-Z relation does actually exist, its most important
consequence is that the distances of objects obtained by SNe~Ia technique 
could have a systematic error, as the SNe~Ia absolute magnitude has not been
accurately estimated.

\begin{figure}[t]
\includegraphics[width=0.45\textwidth,angle=0]{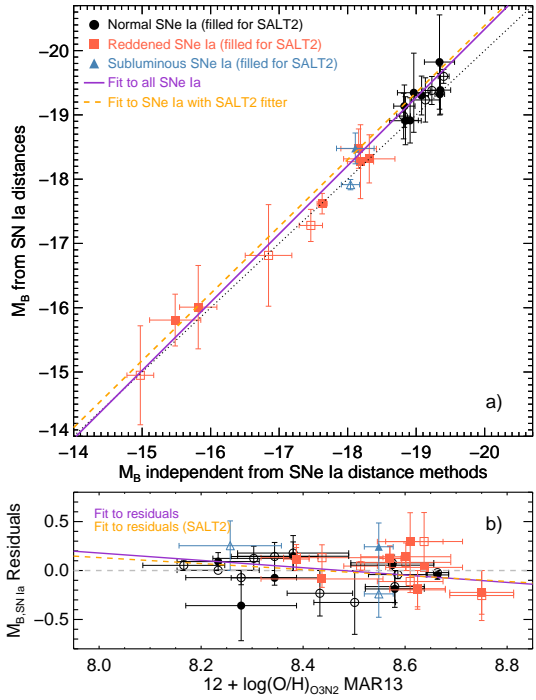}
\caption{a) $M_{B}$ obtained with distances given by the own SN~Ia
technique {\sl vs} $M_{B}$ computed with distances estimated with other independent
methods. The dotted line is the $x=y$
line, while purple solid and dashed orange lines are linear fits to the data. b) $M_{B}$ residuals from upper panel vs. metallicity. }
\label{mb-mb}
\end{figure}

Figure~\ref{mb-mb}a shows both magnitudes $M_{B}$
for our SNe~Ia sample, as obtained from SN~Ia technique distances \citep[taken from NED, prioritizing SALT2 fitter,][]{guy07},
and those derived using SNe~Ia independent methods, e.g.,
Cepheids or Tully-Fisher.
A linear fit to SALT2 data yields:
\begin{equation} M_{B,\,{\rm SN\,Ia}}= 0.46(\pm 1.59) + 1.04 (\pm
0.11) M_{B,\,{\rm SN\,Ia~ind.}},
\end{equation}
(correlation coefficient $r$=0.91). Attending to this figure,
and although the slope is close to the unity, the SN~Ia technique provides higher luminosities (i.e. higher
distances) than the values derived following other methods. {\bf That is, luminosities are
lower than those predicted by SNe Ia techniques}. 
Figure \ref{mb-mb}b shows the $M_{B,SN Ia}$ residuals a a function of oxygen abundance. A linear fit to SALT2 data (orange line) yields:
\begin{equation} \Delta M_{B,\,{\rm SN\,Ia}}=2.57 (\pm
1.95) -0.30(\pm 0.23) {\rm OH_{O3N2}}
\end{equation}
This would have an impact on the HD of cosmological models. Since magnitudes estimated by the color and LC parameters 
provide higher luminosities at high metallicities than they really are, 
some residuals with a positive slope will be induced in the HD when comparing the luminosity of the SN~Ia and the
oxygen abundance (or any other metallicity proxy, as the stellar mass
following the mass-metallicity relation). Such behavior has been in fact 
 observed, since \citet{dandrea11,childress13} and \citet{pan14} found that, once splitting
their SNe~Ia sample in sub-solar and over-solar metallicity, 
those located in high-metallicity hosts are, on average, 0.10-0.12\,mag brighter than those found
in low-metallicity galaxies. Our results explain this fact with a difference of 0.14 $\pm$ 0.10~mag between objects at high and low metallicity host galaxy.
In summary, it seems SNe~I stretch-color-corrected luminosities have a dependence on the properties of their
host galaxies, in particular on the oxygen abundance\footnote{The SN Ia luminosity has a stronger dependence on the gas-phase
metallicity than on the stellar one \citep[see][]{pan14}, probably due to the 
problems of evolutionary synthesis codes, used to determine this last one}, which could artificially increase this
luminosity above the real value.

Therefore, we here suggest to formally
include the metallicity-dependence in the determination of the
distance modulus, $\mu$, as:
\begin{equation} \mu=m_{B}-M_{B}+\alpha \, x_{1}-\beta\,c+\gamma\,{\rm Z},
\end{equation} 
with $\alpha$, $\beta$ and $\gamma$ being the coefficients for the
dependence on stretch, $x_{1}$, color $c$ and metallicity, $Z$, similarly to \citet[][ Eq.2]{lam10} and
\citet{sullivan11}, who consider the stellar mass as an extra
parameter. It should minimize the small but quantifiable
systematic errors induced by the metallicity-dependence of the SNe~Ia 
maximum luminosity.

\section{SUMMARY AND CONCLUSIONS}
\label{conclusions}
\begin{itemize}

\item We estimate oxygen abundances of a sample of 28 local star-forming galaxies hosting
SNe~Ia, in the region where each one exploded, and study the relation with the maximum magnitude $M_B$. Data indicate with a 80\% of chance not to be due to random fluctuation, that most metal-rich galaxies seem to host fainter SNe Ia.

\item These observational data agree with theoretical predictions
from \cite{bravo10}.

\item The existence of such a $M_B$-Z relation would naturally explain the
observational result after correcting for the LC parameters, that
brightest SNe~Ia are usually found in metal-rich or massive galaxies. 
The standard calibration tends to overestimate the maximum luminosities
of SNe~Ia located in metal-rich galaxies. 

\end{itemize}

This relation, as obtained here with star-forming galaxies, may indicate the
metallicity of the progenitor plays a role in the SN Ia luminosity and hence, in the estimated distances. 
It could also be that the host galaxy extinction, not considered in this work, 
correlates with the observed O/H. This variation with O/H would induce systematic errors when using SNe Ia
to derive cosmological distances. 

\acknowledgements
The authors acknowledge the anonymous referee for his/her helpful comments.
This work has been partially supported by MINECO-FEDER-grants
AYA2010-21887-C04-02 and AYA2011-22460.
%
Support for LG is provided by CONICYT through FONDECYT grant 3140566, and by the Chilean's Millennium Science Initiative through grant IC120009, awarded to The Millennium Institute of Astrophysics, MAS.
We thank Eduardo Bravo for their comments to this paper.
M.E.M.-R. thanks the hospitality of both the Australian
Astronomical Observatory in 2013 and the Departamento de
Astronom\'{\i}a of the Universidad de Chile in 2014 during his stays.
Based on observations made with the 4.2m WHT on the island of La Palma by the Isaac Newton Group of
Telescopes in the Spanish observatory of Roque de Los Muchachos of the
Instituto de Astrof\'{\i}sica de Canarias.
This research has made use of the NASA/IPAC Extragalactic Database
(NED), operated by the Jet Propulsion Laboratory, California
Institute of Technology, under contract with the National Aeronautics
and Space Administration.

\end{document}